\documentclass{ws-ijmpb}
\usepackage[12pt]{xy}
\usepackage{graphics}
\usepackage{colordvi}

\begin{document}
\markboth{Magdalena Stobi\'nska, Krzysztof W\'odkiewicz}{Witnessing
  Entanglement of EPR States \\
  with Second-Order Interference}

\title{Witnessing Entanglement of EPR States With \\
  Second-Order Interference}

\author{MAGDALENA STOBI\'NSKA} \address{Instytut Fizyki Teoretycznej,
  Uniwersytet Warszawski, Warszawa 00--681,
  Poland\\magda.stobinska@fuw.edu.pl}

\author{KRZYSZTOF W\'ODKIEWICZ} \address{Instytut Fizyki Teoretycznej,
  Uniwersytet Warszawski, Warszawa 00--681, Poland \\Department of
  Physics and Astronomy, University of New Mexico, Albuquerque,
  NM~87131-1156, USA\\wodkiew@fuw.edu.pl}

\maketitle

\begin{abstract}
  The separability of the continuous-variable EPR state can be tested
  with Hanbury-Brown and Twiss type interference.  The second-order
  visibility of such interference can provide an experimental test of
  entanglement. It is shown that time-resolved interference leads to
  the Hong, Ou and Mandel deep, that provides a signature of quantum
  non-separability for pure and mixed EPR states. A Hanbury-Brown and
  Twiss type witness operator can be constructed to test the quantum
  nature of the EPR entanglement.
\end{abstract}


\section{Introduction}

The famous debate between Einstein, Podolsky, Rosen (EPR) and Bohr
\cite{EPR1935,Bohr1935}, about the nature of quantum correlations of a
bi-partite state has played a key role in the investigation of
entanglement properties of light in modern Quantum Optics, and has
initiated a new branch of physics called Quantum Information. EPR used
in their arguments a~wave function that exhibits a perfect
correlations between positions and momenta of two massive particles
(labeled $a$ and $b$). In the position and momentum representations
the EPR state takes the following forms
\begin{equation}
 \Psi(x_a,x_b) \simeq \delta(x_a-x_b), \quad
\widetilde{\Psi}(p_a,p_b) \simeq
  \delta(p_a+p_b).
\label{EPR}
\end{equation}

Quantum correlations of the entangled EPR state have been implemented
experimentally for massless particles: photons. In the case of light,
the quantum mechanical position and momentum observables are played by
the electric field: phase and amplitude quadratures.

The well known two-mode Gaussian squeezed states of light are physical
realization of the EPR state. Those states lie at the heart of quantum
cryptography \cite{crypt}, quantum information \cite{Braunstein2002}
and quantum teleportation \cite{Furusawa1998}.

There are few efficient ways of producing Gaussian EPR correlated
states. One of them uses a Kerr nonlinear medium in optical fibers to
entangle phase and amplitude quadratures \cite{Silberhorn2001}.
However, the nonlinearity has to be relatively small to achieve a
Gaussian state. The other method uses a beam splitter as a nonlocal
operation that creates entanglement. A typical setup consists of two
initially separable amplitude squeezed beams which interfere at the
50/50 beam splitter.

Recent applications of entangled two-mode squeezed states of light in
quantum information processing have generated a lot of interest in
mixed Gaussian states. Although there are mathematical criterions for
the entangled properties for mixed two-partite Gaussian systems
\cite{Englert2003}, they are not easy experimentally realizable.

It is the purpose of this paper to show that Hanbury-Brown and Twiss
(HBT) interference can reveal the quantum nature of entanglement of
two-particle mixed Gaussian states. We show that it is possible to
construct a quantum witness operator, closely related to the HBT
interference \cite{Stobinska2004}, that probe entanglement of a mixed
EPR Gaussian state. Using Hong, Ou and Mandel (HOM) interferometry we
discuss the entangled properties of time resolved interference of the
EPR state.  We show that the deep in HOM interference can provide a
useful tool to study quantum separability of continuous-variable
Gaussian EPR states.

\section{Mixed EPR states}

A non-degenerated optical parametric amplification, involving two
modes of the radiation field, provides a physical realization of the
EPR state (\ref{EPR}). The quantum state generated in this process has
the following form

\begin{equation}
\label{EPR2} |\Psi\rangle =  \sum_{n=0}^{\infty}\, \sqrt{p_n}\,
|n,n\rangle,
\end{equation}
where $p_n = \bar{n}^{n}/(1+\bar{n})^{n+1}$ is a thermal distribution
with a mean number $\bar{n}$ of photons in each mode.

Using field quadratures eigenstates: $|x_a,x_b\rangle$, we obtain that
the wave function of such a system (\ref{EPR2}) is Gaussian and has
the form

\begin{equation}
\langle x_a,x_b|\Psi\rangle=\frac{1}{\sqrt{\pi}}
e^{-(\bar{n}+\frac{1}{2})(x_a^2+x_b^2) +
2\sqrt{\bar{n}(1+\bar{n})} x_a x_b}\,. \label{NOPA}
\end{equation}
In the limit of $\bar{n} \rightarrow \infty$, the two-mode squeezed
state (\ref{NOPA}) becomes the original EPR state (\ref{EPR}). The
state (\ref{NOPA}) is not entangled only if $\bar{n}=0$.

The simplest mixed generalization of the EPR state involves a Gaussian
density operator $\rho_{ab}$, being a Gaussian operator of the field
modes described by the annihilation and creation operators
$(a,a^{\dagger})$ and $(b,b^{\dagger})$. This Gaussian density
operator of the two modes is fully characterized by its second moment
expectation values of the modes. In the case of a mixed EPR state the
only non-vanishing field correlations are

\begin{equation}
\label{EPRcorr}
 \langle a^{\dagger}a \rangle = \langle b^{\dagger}b \rangle =
 \bar{n}, \ \ \langle ab \rangle = -m_c,
\end{equation}
where $\bar{n}$ is a mean number of photons in each mode, and $m_c$
correspond to the amount of correlation between the two modes. For
$|m_c|= \sqrt{\bar{n}(\bar{n}+1)}$, the EPR state reduces to the pure
state given by (\ref{NOPA}).

As it has been discussed in several papers (see the tutorial
\cite{Englert2003} and references therein), this mixed EPR state is
separable for $\bar{n}>|m_c|$ and its density operator can be
expressed in a sum of product states (Werner separability criterion)

\begin{equation}
\rho_{\mathrm{ab}} = \sum_{i} p_i \, \rho^{i}(a) \otimes
\rho^{i}(b),
\end{equation}
where $\rho^{i}(a)$, $\rho^{i}(b)$ are the density operators of the
two modes and $\sum_{i} p_i =1$, with $0\leq p_i \leq 1$. The mixed
EPR state is entangled and non-separable if

\begin{equation}\label{nonsep}
\bar{n}<|m_c|<\sqrt{\bar{n}(\bar{n}+1)}\,.
\end{equation}

\section{Hanbury-Brown and Twiss Interference with EPR Pairs}

Second and higher orders of coherence of a light beam can reveal its
quantum nature, which cannot be observed in Young-type interference
experiments. Second-order interference involving intensity-intensity
correlations have been first applied by Hanbury-Brown and Twiss in
stellar interferometry \cite{HBT1956}.  In modern Quantum Optics,
second-order interference has been used as powerful tool to study
nonclassical properties of light \cite{ScullyZubairy}.

\begin{figure}[h]
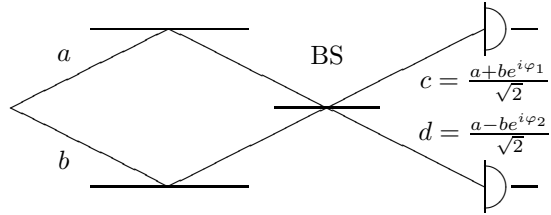

\begin{center}
  \begin{displaymath}
    \xy<0.7cm,0cm>:
    (-1.5,1.5);(1.5,1.5)**@{-},
    (-1.5,-1.5);(1.5,-1.5)**@{-},
    (2,0);(4,0)**@{-}, (3,1)*{\mbox{BS}},
    (-3,0);(0,1.5)**@{-};(6,-1.5)**@{-}, 
    (-3,0);(0,-1.5)**@{-};(6,1.5)**@{-}, 
    (-2,1)*{a}, (-2,-1)*{b},
    (6,2);(6,1)**@{-}, (6,1.5)*\cir(0.5,0.5){r^l},
    (6.5,1.5);(7,1.5)**@{-},
    (6,0.5)*{c=\frac{a+b  e^{i\varphi_1}}{\sqrt{2}}},
    (6,-2);(6,-1)**@{-}, (6,-1.5)*\cir(0.5,0.5){r^l},
    (6.5,-1.5);(7,-1.5)**@{-},
    (6,-0.5)*{d=\frac{a-b e^{i\varphi_2}}{\sqrt{2}}},
   \endxy
  \end{displaymath}
\end{center}
\caption{Hanbury-Brown and Twiss second-order interference.}
\label{Fig:2}
\end{figure}

Hanbury-Brown and Twiss (HBT) interference measures a second-order
normally ordered intensity-intensity correlation function. In
Fig.~\ref{Fig:2} we have depicted a setup for HBT interference that
involves two light beams with annihilation operators $a$ and $b$
interfering at the beam splitter.  A correlation between clicks of two
detectors corresponds to a normally ordered intensity-intensity
correlation function.  The HBT interference exhibits a typical pattern
of the form: $1+v^{(2)}\cos(\varphi_1-\varphi_2)$, where $\varphi_1$
and $\varphi_2$ are phase differences between the beams $a$ and $b$ in
front of the detectors. These phases include geometrical phases and
phases due to the possible action of the beam splitter. In this setup
the two phases can be controlled experimentally. $v^{(2)}$ is a
second-order interference visibility.  For a classical source of light
this visibility is always bounded: $v^{(2)} \le \frac{1}{2}$. This
classical limit is violated for single photons, as it has been shown
in the pioneering experiments performed by Mandel~\cite{Mandel1999}.

We shall apply the HBT setup to study the mixed Gaussian EPR state. At
the detectors ($i=1,2$) the positive-frequency part of electric field
corresponding to modes $c$ and $d$ (normalized to the number of
photons) is as follows
\begin{equation}
\label{eq:bs_transformation}
E^{(+)}(\varphi_i) = \frac{a \pm be^{i\varphi_i}}{\sqrt{2}}\,.
\end{equation}

\noindent
The corresponding field intensity operator at the screen is equal to

\begin{eqnarray}
I(\varphi_i)&=&E^{(-)}(\varphi_i)E^{(+)}(\varphi_i) = \frac{1}{2}(
a^{\dagger}a + b^{\dagger}b \pm b^\dagger a\,e^{-i\varphi_i} \pm
a^\dagger b\,e^{i\varphi_i}).
\end{eqnarray}

From this expression, we obtain that the normally ordered second-order
intensity correlation is

\begin{eqnarray}
\label{eq:HBT_correlation}
 \langle{:}I(\varphi_1)\,I(\varphi_2){:}\rangle
&=& \frac{1}{4}\Big[ \langle :(a^{\dagger}a+b^{\dagger}b)^2:
\rangle - 2\langle a^{\dagger}a\, b^{\dagger}b\rangle
\cos(\varphi_1-\varphi_2)
\nonumber\\
&+& (e^{-i\varphi_1} - e^{-i\varphi_2})\langle
b^{\dagger}(a^{\dagger}a+b^{\dagger}b)\,a\rangle
+ (e^{i\varphi_1} - e^{i\varphi_2})\langle
a^{\dagger}(a^{\dagger}a+b^{\dagger}b)\,b\rangle
\nonumber\\
&-& e^{-i(\varphi_1+\varphi_2)}\langle {b^{\dagger}}^2a^2  \rangle
- e^{i(\varphi_1+\varphi_2)}\langle {a^{\dagger}}^2 b^2 \rangle
\Big]\,.
\end{eqnarray}
For the EPR beams in a mixed state, with mean $\bar{n}$ photons and
correlation $m_c$ described by the relations from Eq.(\ref{EPRcorr}),
the fourth-order field correlations needed in the HBT calculations are
equal to

\begin{equation}
\label{4correlations} \langle a^{\dagger}a^{\dagger} a\,a \rangle
= 2 \bar{n}^2\,,\quad \langle b^{\dagger}b^{\dagger} b\,b \rangle = 2
\bar{n}^2\,,\,\quad
\langle a^{\dagger} a\, b^{\dagger} b \rangle = \bar{n}^2 + |m_c|^2\,.
\end{equation}
As a result the HBT intensity correlation function
(\ref{eq:HBT_correlation}), for a continuous variable mixed EPR state,
has the following form
\begin{equation}
\langle{:}I(\varphi_1)\,I(\varphi_2){:}\rangle =
\frac{1}{2}(3\bar{n}^2 + |m_c|^2) \left\{1-  v^{(2)}
\cos(\varphi_1 -\varphi_2) \right\}\,,
\end{equation}
with the second-order fringe visibility equal to
\begin{equation}
v^{(2)} = \frac{\bar{n}^2 + |m_c|^2}{3\bar{n}^2 + |m_c|^2}\,.
\label{eq:visibility_epr}
\end{equation}

For a separable (classical) output state we have: $\frac{1}{3} \leq
v^{(2)} \leq \frac{1}{2}$. In the case of no correlation in the output
state $|m_c| =0$, the visibility is $v^{(2)} = \frac{1}{3}$, as it
should be for a thermal state. For entangled states this visibility is
quantum i.e., it violates the classical inequality: $v^{(2)}\leq
\frac{1}{2}$.

\begin{figure}[h]
\begin{center}
  \scalebox{0.75}{\includegraphics{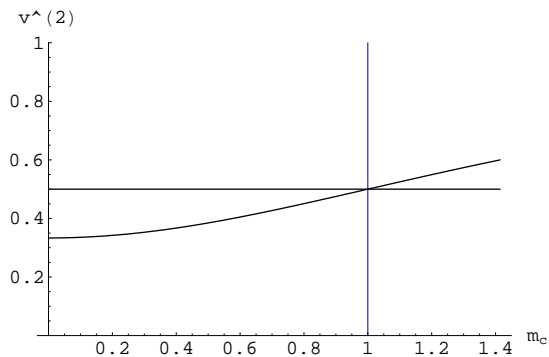}}
\caption{The visibility (\ref{eq:visibility_epr}) evaluated for
  $\bar{n}=1$. The vertical grid line defines a border between
  entangled and separable EPR states. For $|m_c|>1$ the state is
  entangled and the visibility is greater than $\frac{1}{2}$ -
  horizontal grid line.}
\label{Fig:visibility_1}
\end{center}
\end{figure}

\begin{figure}[h]
\begin{center}
  \scalebox{0.75}{\includegraphics{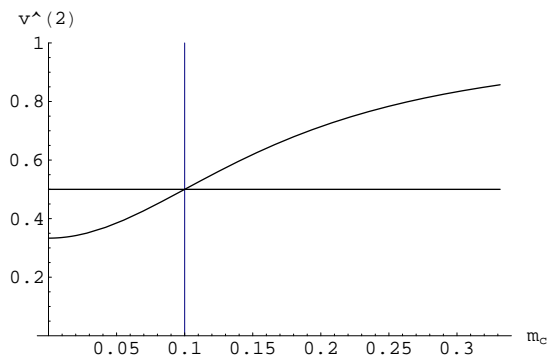}}
\caption{The visibility (\ref{eq:visibility_epr}) evaluated for
  $\bar{n}=0.1$. The vertical grid line defines a border between
  entangled and separable EPR states. For $|m_c|>0.1$ the state is
  entangled and the visibility is greater than $\frac{1}{2}$ -
  horizontal grid line.}
\label{Fig:visibility_2}
\end{center}
\end{figure}

In Fig. \ref{Fig:visibility_1} and Fig. \ref{Fig:visibility_2} we have
depicted the visibility (\ref{eq:visibility_epr}) as a function of the
correlation parameter $m_c$ for two different values of
$\bar{n}=1,0.1$.

Based on the above visibility analysis one can introduce a HBT
witness operator

\begin{equation}
\label{eq:witness}
 \mathcal{W}^{(HBT)}=\frac{1}{2}- \frac{2 a^\dagger  b^\dagger a\, b}
 {\langle :(I_a+I_b)^2: \rangle}.
\end{equation}
The mean value of this operator,
\begin{equation}
\mathrm{Tr}\{\mathcal{W}^{(HBT)}\rho_{\mathrm{AB}}\}=  \frac{\bar{n}^2
-|m_c|^2}{2(3\bar{n}^2 + |m_c|^2)}\,,
\end{equation}
is positive for separable mixed EPR states and negative for entangled
mixed EPR states.

\section{Hong-Ou-Mandel Time-Resolved Interference}

The analysis of second-order interference, can be extended to the case
of photon pulses which duration is long compared to the time
resolution of photodetector. Such single-photon emitters are already
in use \cite{Kuhn2002,Kuhn2003}.

A source produces single photon pulses with a relative delay which
interfere at the beam splitter. Simple analysis of beam splitter's
action on impinging photons shows, that both photons will leave the
same outport of the beam splitter. A relative delay is small enough to
keep the pulses partially overlapping. Even though the overlapping is
not complete, its time duration is longer than the time resolution of
the detector. Therefore, the temporal effects cannot be neglected any
more when detecting pulses and the time delay $\tau$ between detected
pulses at two separate detectors has to taken into account.
Experiments investigating interference for such pulses with different
frequencies have already been performed \cite{Rempe2004,Rempe2003}.

We restrict our description of the single-photon wave packets in the
space-time domain to one-mode fields with electric field operators:
$E^{(+)}_a (t)= \alpha(t)\,a$ and $E^{(+)}_b (t)= \beta(t)\,b$.  The
phases $\varphi_1$ and $\varphi_2$ are included in the mode functions
phases. These mode phases contain an additive stochastic component
related to the random arrival time of the correlated photons at the
beam splitter or detectors.  The time dependent electric fields of the
two-mode output state is described by the beam splitter transformation
(see Eq. (\ref{eq:bs_transformation}))

\begin{equation}
E^{(+)}_c (t) = \frac{E^{(+)}_a (t)+ E^{(+)}_b (t)}{\sqrt{2}}, \quad
E^{(+)}_d (t) = \frac{E^{(+)}_a (t)- E^{(+)}_b (t)}{\sqrt{2}}.
\end{equation}

The joint coincidence of detecting photons from mode $c$ and $d$ at
delayed times $t$ and $t+\tau$ can be obtained directly from the
following second-order temporal coherence function

\begin{equation}
G^{(2)}(t, t+\tau) = \left\langle E^{(-)}_c(t)\, E^{(-)}_d(t+\tau)\,
E^{(+)}_d(t+\tau)\, E^{(+)}_c(t)\right\rangle.
\end{equation}
Using the mode functions, with random phases in the output state, we
note that the only contributing terms to the second-order coherence
function are the following

\begin{eqnarray}
G^{(2)}(t, t+\tau) &=& \frac{1}{4} \Big\{
|\alpha(t)|^2|\alpha(t+\tau)|^2\; \langle a^{\dagger}a^{\dagger}
  a\,a\rangle + |\beta(t)|^2|\beta(t+\tau)|^2\; \langle
  b^{\dagger}b^{\dagger} b\,b\rangle
\nonumber \\
&+& \Big[|\alpha(t)|^2|\beta(t+\tau)|^2
+ |\alpha(t+\tau)|^2|\beta(t)|^2
\nonumber\\
&-&\big(\alpha(t+\tau)\alpha^*(t)\beta(t)\beta^*(t+\tau) + \textrm{c.c.}\big)
\Big]\, \langle a^{\dagger}b^{\dagger} a\, b\rangle\Big\}.
\end{eqnarray}

As an example we shall use in the calculations a special model of
phase fluctuations. We will assume that the mode functions have random
phases typical for a stationary stochastic phase diffusion model. In
this example, the only non-vanishing autocorrelations of the mode
functions are

\begin{equation}
\langle \alpha^*(t)\alpha(t+\tau)\rangle = I_a\,
\exp{\left(-\frac{\tau^2}{\tau_a^2}\right)},\quad
\langle \beta^*(t)\beta(t+\tau)\rangle = I_b\,
\exp{\left(-\frac{\tau^2}{\tau_b^2}\right)}\,.
\label{eq:mode_fun_correlations}
\end{equation}
In these formulas we have assumed that the statistical correlations
are stationary and Gaussian with coherence times $\tau_a$ and
$\tau_b$. The stationary intensities are: $\langle
\alpha^*(t)\alpha(t)\rangle = I_a$ and $\langle
\beta^*(t)\beta(t)\rangle = I_b$.

Applying the above temporal correlations of the mode functions,
combined with the EPR correlations of a mixed state  given by Eq.
(\ref{4correlations}), we obtain that the second-order coherence
function is stationary (dependents only on time~$\tau$) and has
the form

\begin{equation}
G^{(2)}(\tau) = \frac{1}{2} \left\{ \bar{n}^2\, (I_a^2+I_b^2) -
(\bar{n}^2 + |m_c|^2)\, I_aI_b\,
\exp{\left(-\frac{\tau^2}{\tau_a^2}-\frac{\tau^2}{\tau_b^2}\right)}
\right\}\,.
\end{equation}
We shall simplify this formula further assuming that the mode
intensities are equal: $I_a = I_b = I$. Setting $\frac{1}{\tau_a^2} +
\frac{1}{\tau_b^2} = \frac{1}{\tau_c^2}$ the formula above reduces to

\begin{eqnarray}
G^{(2)}(\tau) &=& \frac{I^2}{2}\, (3\bar{n}^2 + |m_c|^2)\,
\left\{1 - v^{(2)}\,  e^{-\frac{\tau^2}{\tau^2_c}} \right\}\,
\end{eqnarray}
with the second-order visibility $v^{(2)}$ equal to
(\ref{eq:visibility_epr}). From this formula we obtain that a joint
coincidence probability to detect a photon at time $t$ and another
photon at time $t+\tau$ is

\begin{equation}
 p(T)= 1 - v^{(2)}\,  \exp{(-T^2) }\,,
\end{equation}
where $T$ is a dimensionless time with $\tau$ expressed in units
of $\tau_c$.

\begin{figure}[h]
\begin{center}
\scalebox{0.75}{\includegraphics{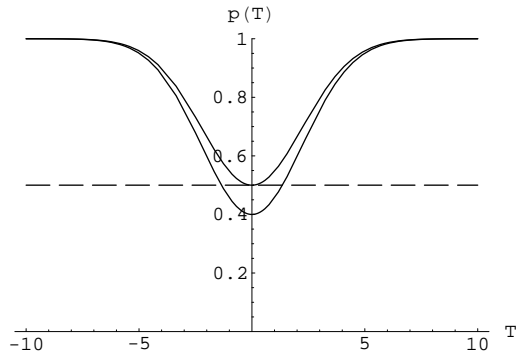}}
\caption{
  The plots of joint coincidence probability $p(T)$ as a~function of
  $T$ for pure (the lower curve) and mixed separable (the upper curve)
  EPR states are depicted.  The minimum value of $p(T)$ for the pure
  state ($\bar{n}=1$ and $|m_c|^2 = 2$) is below $\frac{1}{2}$. For a
  mixed state from the border between entangled and separable states,
  ($\bar{n}=1$ and $|m_c|^2 = 1$), $p(T)$ reaches the value of
  $\frac{1}{2}$ but does not exceed it.}
\label{Fig:ent_pure_1}
\end{center}
\end{figure}

This coincidence probability exhibits a~typical Hong-Ou-Mandel deep
\cite{Hong1987}.  In Fig. \ref{Fig:ent_pure_1}, we have depicted the
joint coincidence probability for a state with $\bar{n}=1$. The upper
curve ($\bar{n}=1$ and $|m_c| =1$) is a border between separable and
nonseparable EPR states.  The lower curve ($\bar{n}=1$ and $|m_c|
=\sqrt{2}$) describes HOM interference of an entangled pure EPR state.
The region between the two curves corresponds to nonseparable mixed
EPR sates. For $\bar{n}=1$ the minimum value the HOM deep for a pure
state is achieved at a zero coincident rate and is equal to
$p_{min}=0.4$.

For the EPR state given by Eq. (\ref{EPR2}), with a small value of
$\bar{n}$, one can get a significant increase of the the deep. For a
weak two-mode pure squeezed state $(\bar{n} \ll 1)$ we can approximate
the formula (\ref{EPR2}) by

\begin{equation}
|\Psi\rangle \sim  \sqrt{p_0}\, |0,0\rangle +\sqrt{p_n}\,
|1,1\rangle\,.
\end{equation}
We see that such a weak pure squeezed state is the same as the one
discussed in the single photon experiment originally performed by HOM
\cite{Hong1987}.

\begin{figure}[h]
\begin{center}
  \scalebox{0.75}{\includegraphics{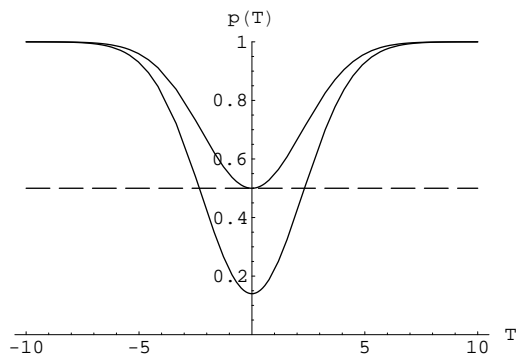}}
  \caption{Similarly, as in Fig.~\ref{Fig:ent_pure_1}, for a~state with
    $\bar{n}=0.1$, probability $p(T)$ exceeds the value of
    $\frac{1}{2}$ for a~pure state ($|m_c|^2 = 0.11$, the lower
    curve). In this case the minimum value is smaller than for
    $\bar{n} = 1$ and reaches $0.14$. For the border state, between
    entangled and separable states ($|m_c|^2 = 0.01$, the upper
    curve), the minimum value of $p(T)$ is again equal to
    $\frac{1}{2}$.}
\label{Fig:ent_pure_01}
\end{center}
\end{figure}

Fig. \ref{Fig:ent_pure_01} illustrates the HOM deep for a state with
$\bar{n}=0.1$. The upper curve ($\bar{n}=0.1$ and $|m_c| =0.1$) is a
border between separable and nonseparable EPR states.  The lower curve
($\bar{n}=0.1$ and $|m_c| =\sqrt{0.11}$) describes HOM interference of
an entangled pure EPR state approximated by the state exhibited above.
The region between the two curves corresponds to nonseparable mixed
EPR sates. For $\bar{n}=0.1$ the minimum value the HOM deep for a pure
state is achieved at a zero coincident rate and is equal to
$p_{min}=0.14$.

\section{Conclusions}

In this paper we have shown that the Hanbury-Brown and Twiss type
interference can be used as an experimental test of quantum
separability for continuous variable EPR states. One can associate
with this second-order interference a HBT witness operator quantifying
entanglement for pure and mixed Gaussian EPR states.  We have shown
that time-resolved coincidence rate of joint photon detection exhibits
the Hong, Ou and Mandel deep, that provides a signature of quantum
nonseparability for pure and mixed EPR states.

\section*{Acknowledgments}

This work was partially supported by a~KBN grant No.
PBZ-Min-008/P03/03.

\end{document}